\documentclass{article}
\usepackage{graphicx}
\usepackage{hyperref}
\usepackage[authoryear]{natbib}

\usepackage{threeparttable}
\begin{document}
\title{Discovery of a young subfamily of the (221)~Eos asteroid family}
\author{Georgios Tsirvoulis\thanks{E-mail: gtsirvoulis@aob.rs}\thanks{Preprint Submitted to MNRAS}\\
Department of Astronomy, Faculty of Mathematics,\\
 University of Belgrade, \\
Studentski trg 16, 11000 Belgrade, Serbia}
\maketitle

\begin{abstract}
In this work we report the discovery of a young cluster of asteroids that originated by the breakup of an asteroid member of the (221)~Eos family. By applying the Hierarchical clustering method to the catalog of proper elements we have identified 26 members of this new small group of asteroids. We have established that the statistical significance of this cluster is $>99\%$, therefore it corresponds to a real asteroid family, named the (633)~Zelima cluster, after its lowest numbered member. The orbits of its members are dynamically stable, a fact that enabled us to use the backward integration method, in two variants to identify potential interlopers and estimate its age.  Applying it first on the orbits of the nominal family members  we identified three asteroids as interlopers. Then we applied it on a set of statistically equivalent clones of each member to determine the age of the cluster, with a result of $2.9\pm0.2$~Myrs. 
\end{abstract}

\section{Introduction}

Asteroid families, groups of asteroids that share common dynamical and physical characteristics, are believed to be the outcomes of collisions between asteroids in the Main Belt \citep{2002aste.book..619Z}. Since the pioneering work of \citet{1918AJ.....31..185H}, a century ago, the study of asteroid families has been a very active field, and the primary source of our knowledge on the collisional history of the Main Belt \citep{2005Icar..175..111B,2009P&SS...57..173C}. However, since asteroid families evolve over time due to various phenomena such as chaotic evolution of their orbits \citep{ 1994Natur.370...40M,2002Icar..157..155N}, the semi-major drift induced by the Yarkovsky effect \citep{1999Sci...283.1507F,2006Icar..182..118V}, further collisional evolution \citep{1999Icar..142...63M,2002Icar..156..191D}, close encounters \citep{2003Icar..162..308C,2010CeMDA.107...35N} and secular resonances  with massive asteroids \citep{2015ApJ...807L...5N,2016Icar..280..300T}, the older a family is, the less likely it is for us to reconstruct its collisional birth. Therefore, young asteroid families which have not evolved significantly allow us to extract information on the physical processes of colliding asteroids in a more direct manner. 

Several such young asteroid families and small clusters have been identified over the past several years \citep{2003ApJ...591..486N,2006Sci...312.1490N,2006AJ....132.1950N,2009Icar..204..580P,2010MNRAS.407.1477N,2011AJ....142...26V,2012MNRAS.425..338N,2014Icar..231..300N,2018Icar..304..110P}. Of particular interest are the cases concerning young clusters that are sub-families of known large, old asteroid families, such as the (832)~Karin cluster, a sub-family of (158)~Koronis \citep{2002Natur.417..720N}, and (656)~Beagle, a sub-family of (24)~Themis \citep{2008ApJ...679L.143N}. The existence of such young sub-families can help constrain the timescale of the evolution of the parent, large family due to collisional disruptions, and in consequence of the whole Main Belt. Another important aspect of young sub-families is the fact that their members have suffered less alterations of their surfaces due to space weathering effects \citep{2009Natur.458..993V} compared to the members of the original family \citep{2016Icar..269....1F}. This  can be of great importance in the study of both the physical processes of space weathering themselves, and of the physical properties of the parent family via the freshly exposed material.

We have discovered a new example of a very old large family having a sub-family of much younger age. The asteroid family of (221)~Eos has been estimated to be about $1.5\,Gyrs$ old \citep{2015Icar..257..275S,2017Icar..288..240M,2006Icar..182...92V,2013Icar..223..844B}, and among its family members we found a young cluster of asteroids only a few million years old. The parent body of the (221)~Eos family is believed to have been partially differentiated \citep{2005A&A...442..727M}, so the discovery of this young cluster might be of great interest. The cluster members, which should be similar compositionally as they originate from a single fragment of the original disruption, could provide a fresh view of the interior of the Eos parent body.

\section{Identification of the new family}

Discovering intentionally a very small young family within one of the densest regions of the Main Belt, as is the region covered by the Eos family, is a very challenging undertaking, with few chances of success. One has to scrutinize every corner of the three dimensional region in question for any abnormalities in the distribution of asteroids, and then judge whether there is any statistically significant group that could be an actual asteroid family, originating from the disruption of a single parent body. Much of this intense labor is eliminated though when the factor of luck manifests; certainly not a rare occasion in science, with the current work also being a beneficiary. As long as the statistical tests hold, nature does not care if we make discoveries by systematic search or by chance. 

\begin{figure}
  \includegraphics[width=0.99\columnwidth]{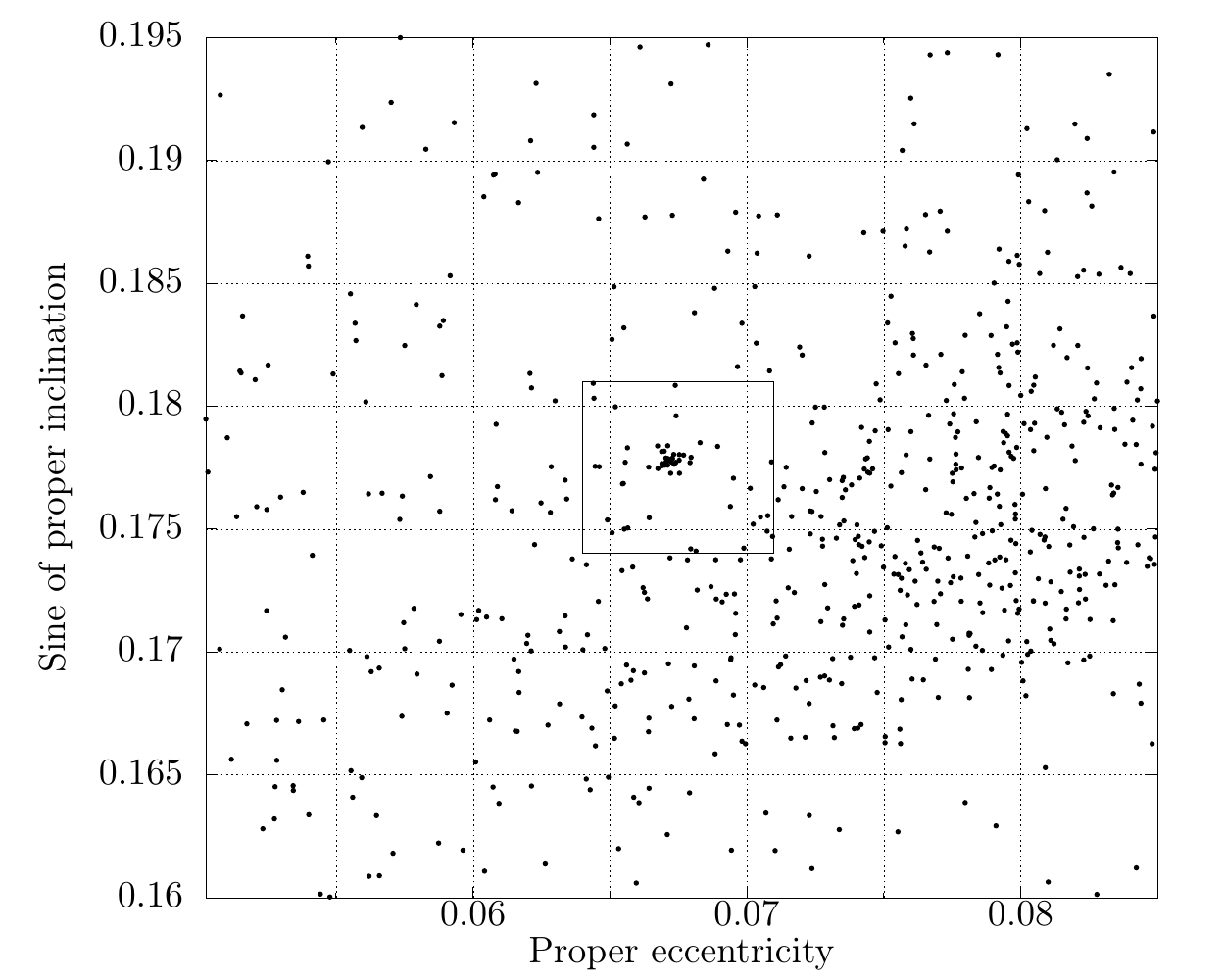}
    \caption{Distribution of asteroids in the $(e_p,\sin{i_p})$ orbital planes, at a sub-region of the Eos family. There is an apparent over-density of asteroids (highlighted with the box), hinting the existence of a young cluster.}
    \label{fig:cluster}
\end{figure}

When looking at the distribution of asteroids in the proper elements space in a small region within the Eos asteroid family, a clustering of a small number of asteroids is apparent, especially in the proper eccentricity versus the sine of proper inclination projection $(e_p,\sin{i_p})$, as seen in \autoref{fig:cluster}.

In order to identify the members of this new family, the hierarchical clustering method (HCM) as proposed by \citet{Zappala1990} was used. Given a set of proper elements and a starting asteroid, the HCM checks whether the closest neighbor is close enough to be linked to it, forming a cluster. The metric used to determine the distance between two asteroids in the proper elements space, is the one proposed by \citet{Zappala1994}:

\begin{equation}
d_c=na_p \sqrt{\frac{5}{4}\left( \frac{\delta a_p}{a_p}\right)^2+2(\delta e_p)^2+2(\delta \sin{i_p})^2}
\end{equation}
where $n$ is the mean motion of an asteroid at a proper semi-major axis $a_p$ and $\delta$ denotes the difference in the value of the corresponding proper element between the two asteroids under examination. The two asteroids are considered to belong to the same cluster if their mutual distance is shorter than a preselected value, called the distance cut-off.

For a given distance cut-off, the process is iterated until no other asteroids can be linked to the cluster and returns the list of asteroids that successfully linked into a cluster. To determine the membership of an asteroid family, the method is repeated for increasing values of the cut-off, yielding the number of associated asteroids as a function of the distance cut-off. If this function presents a plateau, i.e. a range in distance cut-off with no significant increase in the membership, it means that the cluster of asteroids up to that point is relatively isolated from the rest of the asteroid population, signifying a real asteroid family \citep{2005Icar..173..132N}.

\begin{figure}
  \includegraphics[width=0.99\columnwidth]{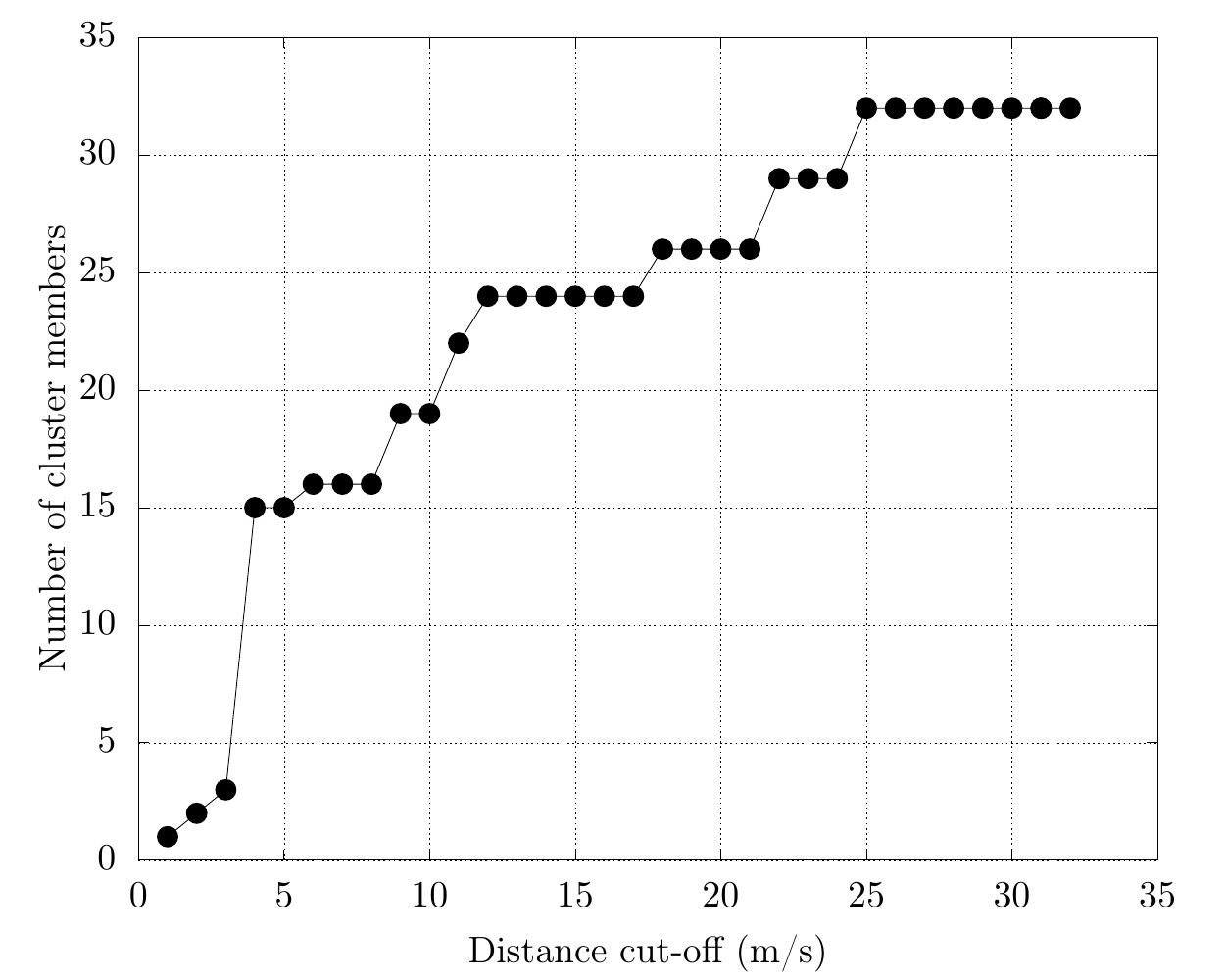}
    \caption{The number of asteroids associated to the (633)~Zelima cluster as a function of the distance cut-off.}
    \label{fig:cutoff}
\end{figure}

From the catalog of synthetic proper elements \citep{Knezevic2000,Knezevic2003} obtained from the AstDyS service\footnote{available at: http://hamilton.dm.unipi.it/astdys/index.php, accessed in June 2017}, all asteroids within the ranges:
\begin{equation}
 2.96<a_p<3.15~\textrm{AU},\\ 
 0.02<e_p<0.13, \\ 
 0.15<\sin{i_p}<0.21
 \end{equation} 
i.e. in the region around the (221)~Eos family, were extracted and used as the initial set. The choice of the starting body is not significant in this case, as long as an asteroid close to the center of the cluster is selected. Due to the apparently tight clustering of asteroids in the small family studied here, very small values of the starting distance cut-off as well as the increasing step of $1~m\, s^{-1}$ was chosen. The result of this process is shown in \autoref{fig:cutoff}; for cutoff greater than $4~m\, s^{-1}$, a value where the core of the cluster is linked together, the number of asteroids associated increases steadily with increasing distance cut-off, until a relatively constant number of asteroids is reached for values between $12 - 17~m\, s^{-1}$, followed by another increase and a second plateau at $25-32~m\, s^{-1}$, after which point the entirety of the population in our catalog is linked together.

Despite the whole procedure being mathematically rigorous up to this point, the choice of the distance cut-off is ultimately subjective, and it is rarely the case that a single value is obviously the correct one. The correlation of the number of asteroids with the distance cut-off needs to be evaluated together with the shape of the growing cluster in the proper elements space to make a choice for the nominal membership which should represent the statistical properties of the cluster despite small deviations. In this case the nominal membership was adopted as the one derived at to $20~m\, s^{-1}$, as at this value the membership corresponds exactly to the isolated cluster as appearing in the $(e_p,\sin{i_p})$ plane. This yields a family membership of 26 asteroids, with the lowest numbered one being (633)~Zelima.

Finding a number of asteroids that can be linked by HCM to a single cluster does not necessarily mean that this cluster corresponds to a real asteroid family, even more so when that group of asteroids is so small. A simple test of the statistical significance of the cluster is necessary to address this question. As this is a subfamily within the Eos family population, the first step was to extract from the proper elements catalog the members of the latter. Then 1000 fictitious clones of the entire Eos family were created by randomly distributing the same number of asteroids within the same orbital volume occupied by the family members.
\footnotetext{Obtained from the Asteroid Families Portal (AFP) at:\\ http://asteroids.matf.bg.ac.rs/fam/index.php}\footnote{As the actual family does not uniformly fill the cubic volume within which it resides, if we selected the actual borders of the family for the box, the mean number density would be smaller than it actually is. To avoid this the dimensions of the box were selected smaller than the size of the family, to the values where the bulk of the Eos family is enclosed, but with the total number of asteroids preserved. This may give an increased mean number density compared to the correct one, but it only means that the statistical significance test is more strict.} Then the fictitious populations were searched by HCM for any clusterings of 26 or more asteroids at a distance cut-off of $20~m\, s^{-1}$, with negative outcome. This leads to the conclusion that a clustering of asteroids with this number of members at such a tight configuration can not be the outcome of chance, but it has to be the outcome of a real collisional breakup. It is important to note that we have only proven the statistical significance of the cluster as a whole. The statistical nature of the HCM does not allow for any conclusions on whether any individual asteroid identified as a member has indeed originated from the breakup event that formed the cluster, or it is an interloper. We will present in the next section a method of identifying interlopers among the identified family members.  

\section{Age of the cluster}

As we have shown, the  Zelima cluster is very compact in the proper elements space,
and the threshold distance needed to identify its members is very small. These
suggest that the cluster itself is a young one, with an age of the order of
a few million years. The age determination method which is most suitable for
young asteroid families and clusters is the Backward Integration Method (BIM) \citep{2002Natur.417..720N}. The BIM is based on the idea that due to the anticipated
young age of a given family or cluster, the orbits of its members still carry
information of the formation event. Due to the chaotic nature of the Solar System in most cases it is impossible to preserve orbital properties and backtrace every step of the evolution. However in the cases of very young clusters, the time passed since the formation event was not sufficient to allow perturbations  to sufficiently mix their orbital elements. Therefore it is possible to directly
follow the history of the cluster by numerically integrating the orbits of its
members.

Immediately after the creation of the cluster, from the disruption of a parent body, the orbits of the fragments are determined by the ejection velocities
through the equations of Gauss, and are almost identical. Planetary perturbations
and non-gravitational effects, such as the Yarkovsky effect, acting on the cluster
members, cause their orbital elements to diverge over time. Using the BIM we
are able to find the age of the cluster by finding the point in the
past where the orbital elements of the member asteroids, and more specifically the secular angles, i.e. the longitude of the ascending node ($\Omega$) and the argument of perihelion ($\omega$) converge, which corresponds to the formation time. This method has been used frequently in the past to determine the ages of various young families and clusters,
such as the Karin cluster \citep{2002Natur.417..720N}, Veritas family \citep{2003ApJ...591..486N}, the Datura cluster \citep{2006Sci...312.1490N}, Theobalda family \citep{2010MNRAS.407.1477N}, the Lorre cluster \citep{2012MNRAS.425..338N}, the Gibbs cluster \citep{2014Icar..231..300N} and the Schulhof family \citep{2016AJ....151...56V}. 

Another important benefit of the BIM is the identification of potential interlopers among the family members as found by HCM (see e.g. \citet{2012MNRAS.424.1432N}). Any particular asteroid that is included in the membership but does not really belong to the collisional family should in principle have a different formation time. As such, its orbital elements  (the secular angles more specifically) should not converge with those of the true family members at the time of the cluster formation, which is the essence of the  method we used for interloper identification for the  Zelima cluster.

\subsection{Orbital evolution of the cluster members}

\begin{figure}
  \includegraphics[width=0.99\columnwidth]{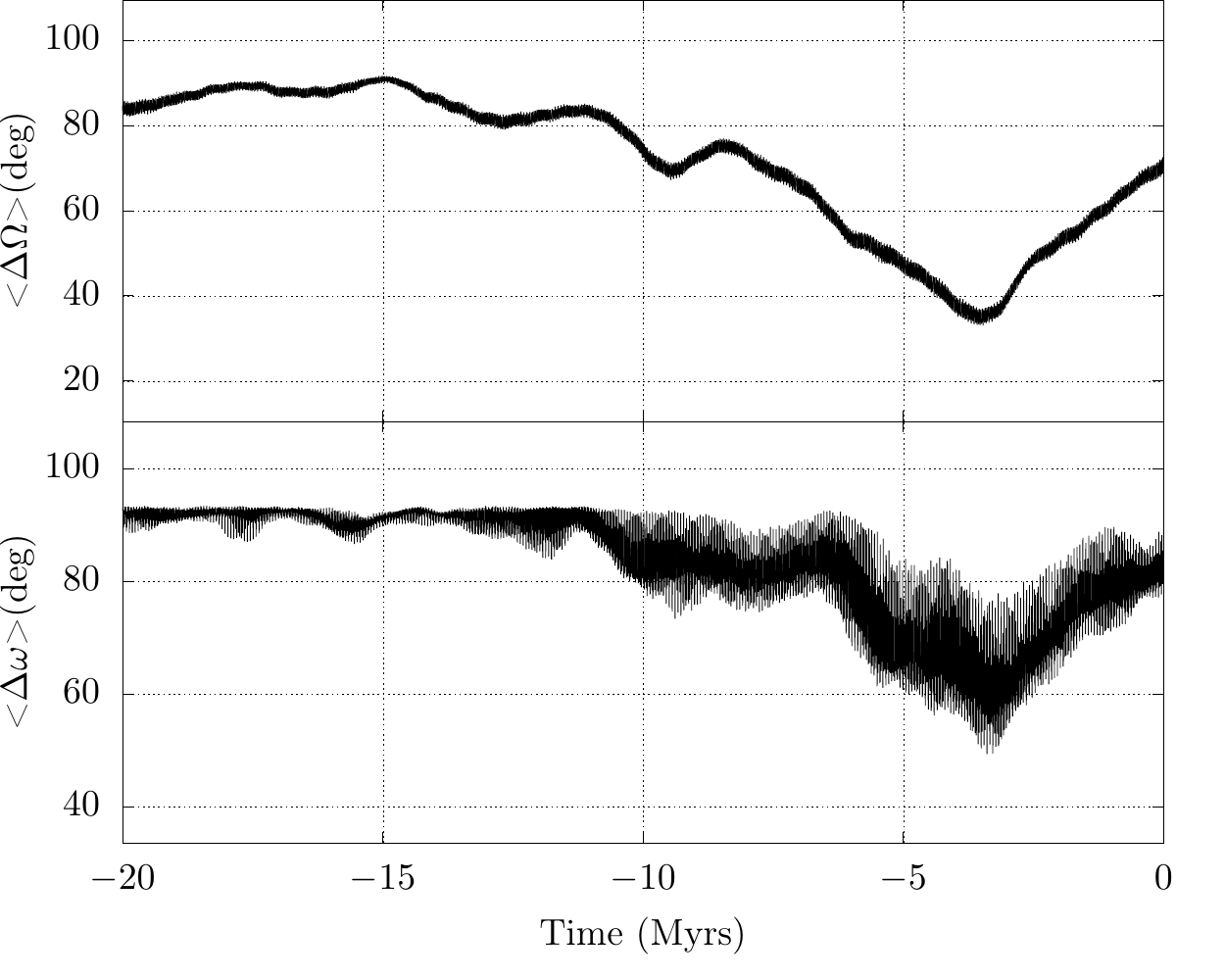}
    \caption{Evolution of the mean differences in the secular angles $\Omega$ (top) and $\omega$ (bottom) of the nominal Zelima cluster members.}
    \label{fig:bim1}
\end{figure}

For the BIM to be reliable, the cluster must be relatively young (up
to about 10 Myrs) and dynamically stable. Both of these conditions are satisfied
by the Zelima cluster, due to the tight packing of its members in the proper
elements space and their long Lyapunov times. The first step was to integrate backwards
in time the orbits of the nominal members of the cluster for 10 Myrs, using a dynamical
model which includes the four outer planets, from Jupiter to Neptune, using
the ORBIT9 integrator \citep{1988CeMec..43....1M}.To account for the
indirect effect of the inner planets, their masses are added to the mass of the Sun
and a barycentric correction is applied to the initial conditions. After retrieving the evolution
of each member's orbital elements, we calculate the mean differences in their
two secular angles, the longitude of the ascending node ($\Omega$), and argument of
perihelion ($\omega$), over timesteps of 500 years, as shown in \autoref{fig:bim1}. The results show
convergence of both angles at about $3.5$ Myrs in the past, within 35$^{o}$ for $ <\Delta\Omega>$
and within 55$^{o}$ for $ <\Delta\omega>$ \footnote{The convergence is worse for $ <\Delta\omega>$ because the corresponding secular frequency at the region of the cluster ($\dot{\omega}=\dot{\varpi}-\dot{\Omega}=g-s\approx145.5''yr^{-1}$) is larger than that for  $ <\Delta\Omega>$ ($\dot{\Omega}=s\approx-71.5''yr^{-1}$), leading to the faster dispersion of the arguments of perihelion}.
Even though both secular angles seem to converge at a single point in the past relatively well, one would expect to have even stronger convergence, i.e. deeper minima. Although in the ideal case one might expect convergence to within one degree, assuming a typical initial dispersion of the fragments corresponding to an escape velocity of the order of $25km\, s^{-1}$, there are a number of reasons  why this is not the case: i) Even though the orbits of most of the cluster members have long Lyapunov times, it is possible that they have spent a period in the past in more chaotic orbits, ii) the Yarkovsky effect, which is not included in the dynamical model, acting upon the members of the cluster, most of which are smaller than 5~km in diameter, can effectively change their semi-major axis even within such a short time, and consequently alter their secular frequencies as the gradients of the secular frequencies  are relatively large in the region of the cluster ($ds/da \approx -70''yr^{-1}AU^{-1}$, $d\omega/da=dg/da-ds/da \approx 200''yr^{-1}AU^{-1}$). iii) The presence of potential interlopers worsens the convergence of the secular angles, as mentioned above, since any interloper essentially adds random terms to the sum of mutual differences in the secular angles when calculating the mean differences.

\begin{figure}
  \includegraphics[width=0.99\columnwidth]{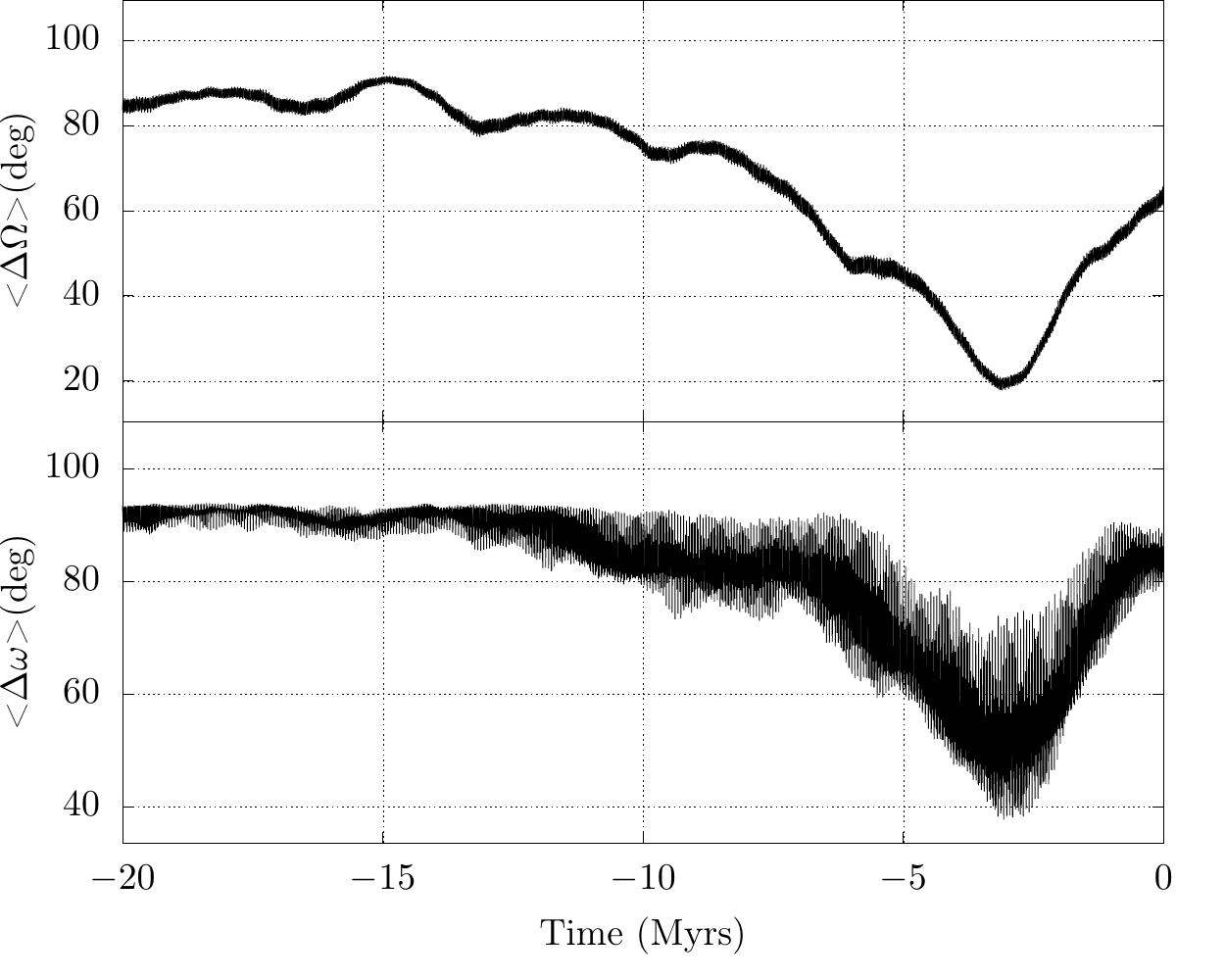}
    \caption{Evolution of the mean differences in the secular angles $\Omega$ (top) and $\omega$ (bottom) of the Zelima cluster members after the three identified interlopers have been removed.}
    \label{fig:bim2}
\end{figure}

In order to identify potential interlopers and remove them from the family membership, we followed a simple yet effective method: Having the numerical integrations at hand we repeated 26 times the calculation of the mean differences in the secular angles at each timestep, each time excluding from the members list one asteroid, leaving only 25. We then compare the resulting plot to that of all 26 members (\autoref{fig:bim1}), and we looked for any improvement in the convergence of the longitude of the ascending node ($\Omega$), as it is much clearer to judge. In order to signify an asteroid as an interloper we demanded that both the mean difference and its standard deviation improve (decrease) by more than 5$^o$ at the time of convergence if that asteroid is excluded from the calculation. In this way we identified three interlopers, namely asteroids (1881), (25583) and (460267). The final result, that is the mean differences in the secular angles as a function of time of the 23 remaining family members, is shown in \autoref{fig:bim2} which features an improved convergence to within about 20$^o$ for $ <\Delta\Omega>$ and about 40$^o$ for $ <\Delta\omega>$, and at an earlier time, at about $3$ Myrs in the past. Let us mention that for the 23 other asteroids there was no measurable improvement at all, of neither the mean difference nor the standard deviation, which means those are indeed real members, at least as far as this method suggests.

\subsubsection{Orbital and Yarkovsky clones}

The reliability of the BIM applied to the orbits of the nominal members of
families or clusters suffers from two major sources of errors. The first one is the
non-zero uncertainty in the orbital elements of their members, and the second
is the secular evolution of their semi-major axis due to the Yarkovsky thermal
force. To avoid the error in the estimated age introduced
by those causes, we use a statistical implementation of the BIM, as originally
proposed by \citet{2006AJ....132.1950N}. This method has also been used
in the past to determine the ages of young clusters, such as the Schulhof family
\citep{2011AJ....142...26V}, the Lorre cluster \citep{2012MNRAS.425..338N} and the Gibbs cluster \citep{2014Icar..231..300N}.

This method uses a set of statistically equivalent orbital and Yarkovsky
clones for each nominal member of the cluster. In detail, we generate 10 orbital clones for each member with orbital elements randomly chosen within the $3\sigma$ uncertainties of its nominal osculating orbital
elements, assuming Gaussian distributions. 

Next, in order to emulate the Yarkovsky thermal force \citep{1998A&A...335.1093V,1999A&A...344..362V}, for each orbital clone we calculate the maximum expected drift rate in semi-major axis according to its diameter, $(da/dt)_{max}(D)$: we first determine the maximum drift rate for a $D=1km$ sized asteroid assuming the following physical parameters: bulk density $\rho =2500 kg\, m^{-3}$, surface density $\rho_s = 1500 kg\, m^{-3}$, thermal
conductivity $K = 0.001 W\, m^{-1}K^{-1}$, specific thermal capacity
$C = 680 J\, kg^{-1}K^{-1}$,  geometric albedo $p_{v}=0.2$, infrared emissivity
$\epsilon = 0.9$, i.e. all typical values for regolith covered basaltic
asteroids \citep{2013Icar..223..844B}. The value obtained is: $(da/dt)_{max}(1km) = 3.4\cdot 10^{-4}~AU\, Myr^{-1}$. Then for an asteroid of diameter D we have $(da/dt)_{max}(D)=(da/dt)_{max}(1km)/D$, as the Yarkovsky effect scales inversely proportionally with the diameter. Finally for each orbital clone we generate 10 Yarkovsky clones, and to each we assign a drift rate randomly selected in the interval $\pm(da/dt)_{max}(D)$ to account for the possible orientations of the spin axis.  In total, 100 statistically equivalent clones are generated for each of the 23 nominal members of the cluster, after the removal of interlopers.
Next we integrate numerically the orbits of all clones backwards in time for
5 Myrs using the ORBIT9 integrator, using the same dynamical model as for the nominal members, i.e. containing only the four giant planets.

\begin{figure}
  \includegraphics[width=0.99\columnwidth]{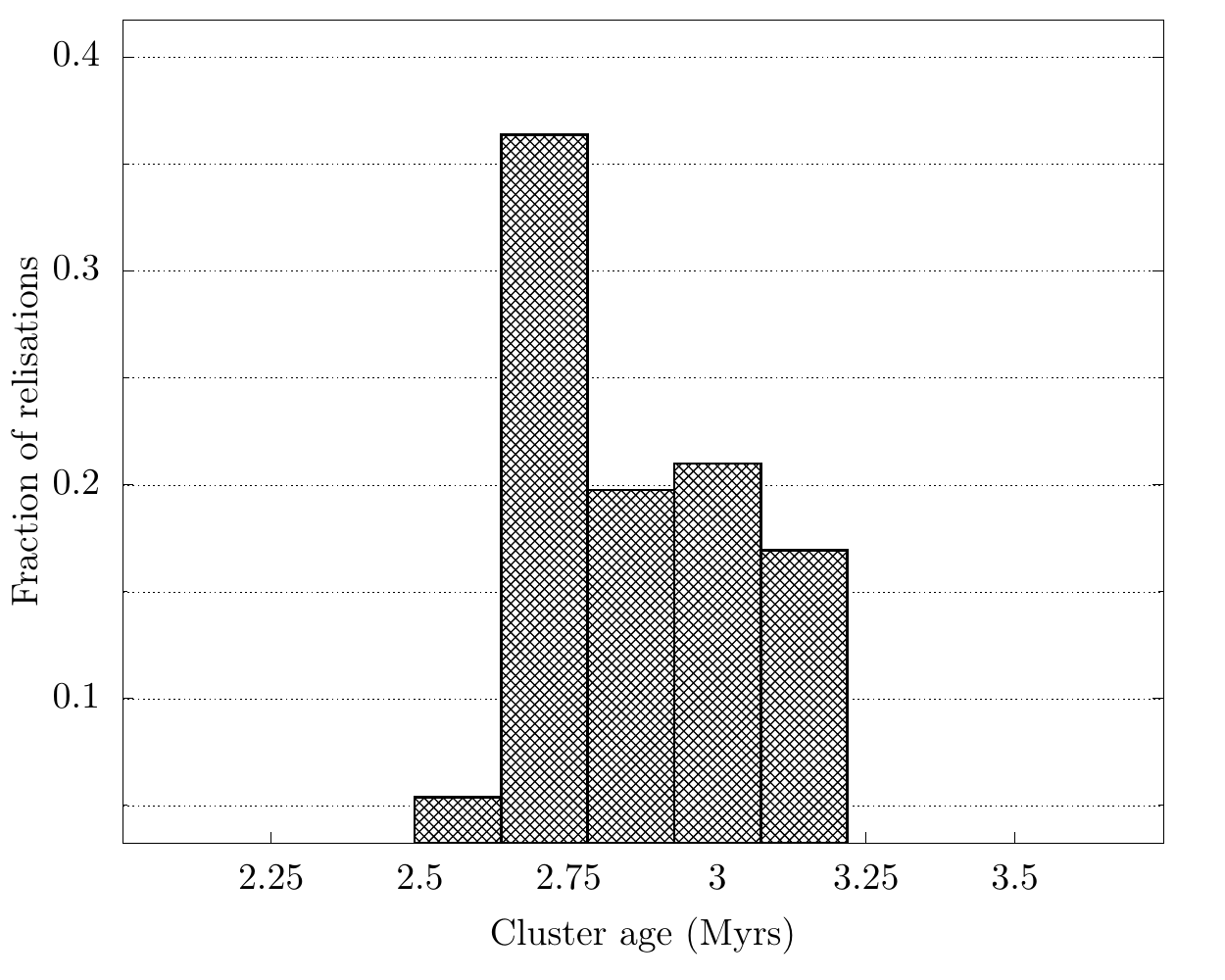}
    \caption{Histogram of the minima of the evaluated \autoref{eq:fun}, revealing the calculated age of the cluster.}
    \label{fig:hist}
\end{figure}

Having obtained the evolution of each clone's orbital elements, we randomly
select one clone for each member and determine the age as the minimum of the
function:
\begin{equation}
\Delta V = n a \sqrt{(sin(i) \Delta \Omega)^{2} + 0.5(e \Delta \omega)^{2}}
\label{eq:fun}
\end{equation}
where $na = 17.14\,km\, s^{-1}$ is the mean orbital speed of the cluster members, $\Delta\Omega$
and $\Delta\omega$ are the dispersions in the longitude of the ascending node and argument
of perihelion respectively, among the 23 clones (e.g.: ($\Delta \Omega)^{2} = (\sum_{ij} (\delta\Omega)^{2}) / N_{pairs}$ ).
We calculate the minimum of this function for $3\cdot 10^6$ random combinations of
clones, which is a sufficiently large statistical sample. In all cases the minimum of the function was less than 3 $m s^{-1}$ with an average value of 2 $m s^{-1}$, which indicates good convergence for all the combinations, bearing in mind that the convergence criterion proposed by \citet{2011AJ....142...26V} is $\Delta V_{min}<5\,m s^{-1}$. \autoref{fig:hist} shows the distribution of the minima of this function, and the resulting age of the cluster was derived to be $2.9\pm0.2$~Myrs.

\section{Dynamical and physical properties}

\begin{figure}
  \includegraphics[width=0.99\columnwidth]{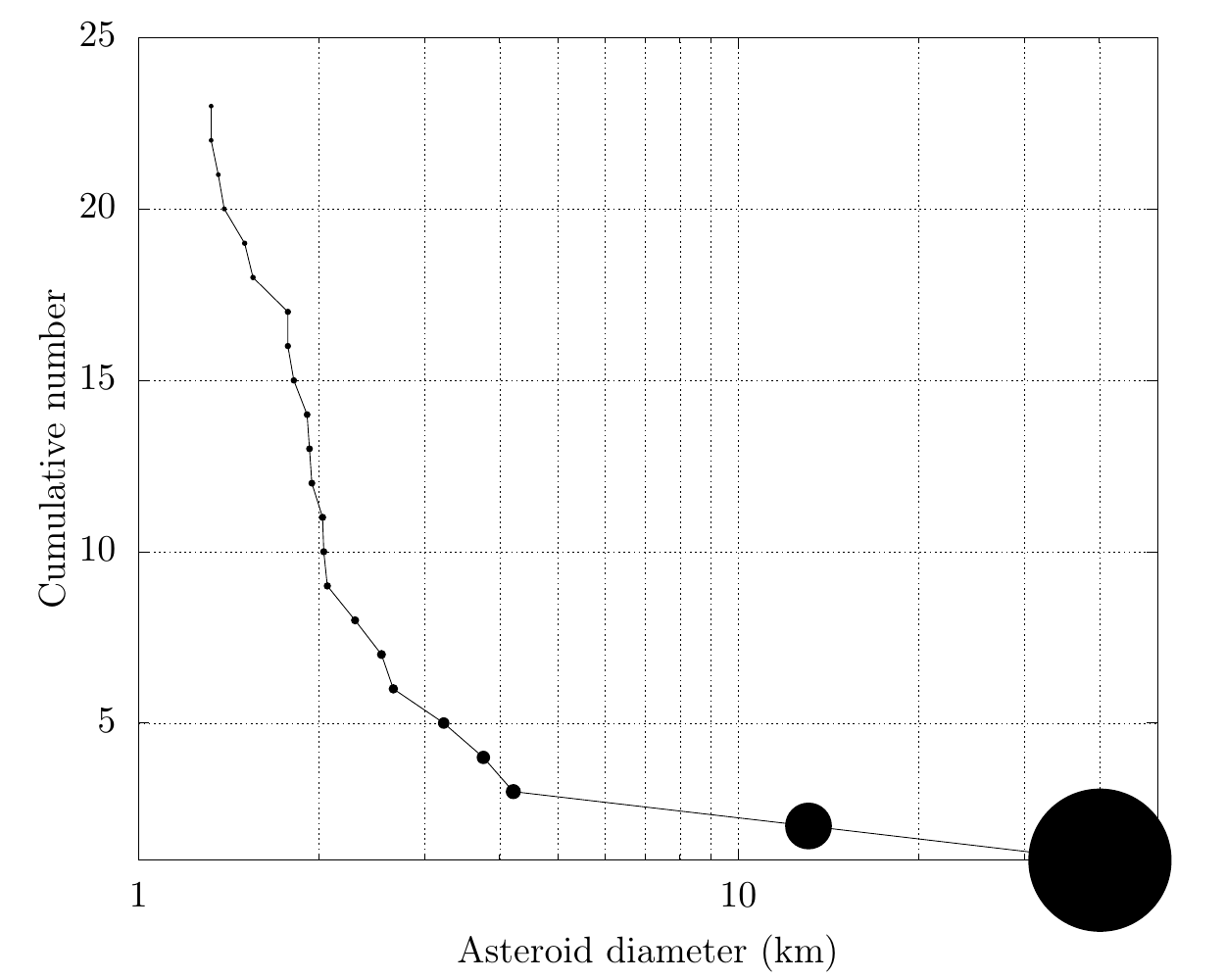}
    \caption{ The cumulative size-frequency distribution of the (633)~Zelima cluster members. }
    \label{fig:cumul}
\end{figure}

We will now present some important properties of the (633)~Zelima cluster and its dynamical environment. In \autoref{tab:example_table} we present the final list of the family members after the three interlopers have been removed. The Wide-field Infrared Survey Explorer (WISE) \citep{2011ApJ...741...68M} provides diameters and albedos for the four largest cluster members. For the rest of the asteroids we calculated their diameter from their absolute magnitudes, assuming a geometric albedo of $p_v=0.2$.  \autoref{fig:cumul} shows the cumulative size-frequency distribution of the family members. We notice that apart from a single asteroid, namely (6733), the largest remnant of the collision is more than one order of magnitude larger than all of the other asteroids, and that the distribution of the latter is rather steep. This suggests that the collision that formed the cluster was a cratering event. We estimated the size of the parent body, assuming it had a spherical shape, by summing up the volumes of all known fragments into a single object. This yielded a diameter for the parent body of  $D_{pb}\approx40.66km$. The ratio of the sizes of the largest remnant, (633)~Zelima, to that of the parent body turns out to be greater than 0.98.

\begin{figure}
  \includegraphics[width=0.54\columnwidth]{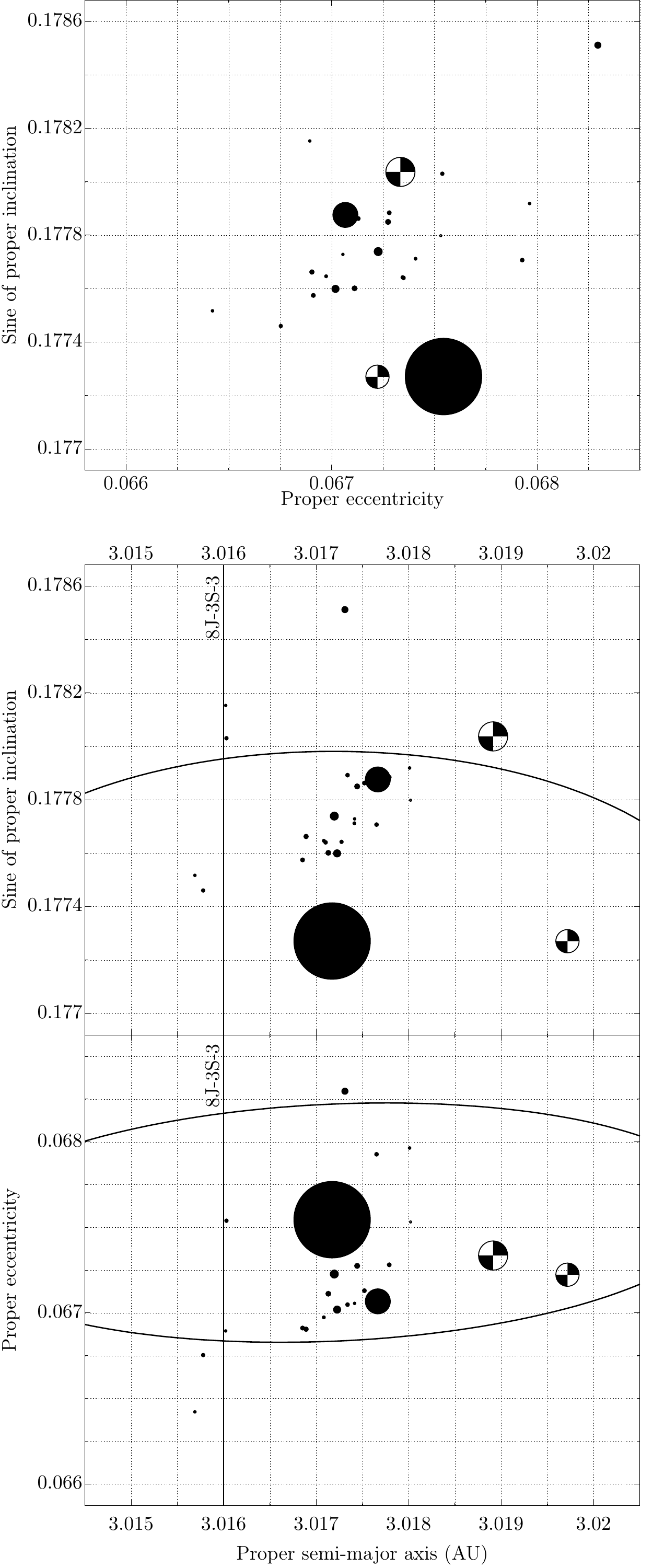}
    \caption{Distribution the Zelima cluster members in the $(e_p,\sin{i_p})$ (top), $(a_p,\sin{i_p})$ (middle) and $(a_p,e_p)$ (bottom) orbital planes. The sizes of the points are proportional to the diameter of each asteroid. The patterned points indicate the identified interlopers. The ellipses in the bottom two panels represent equivelocity curves corresponding to a velocity change of $\Delta v=12ms^{-1}$ from the barycenter of the cluster. The vertical lines show the approximate location of the 8J-3S-3 mean motion resonance (see e.g. \citet{GALLARDO2014273}).}
    \label{fig:cluster2}
\end{figure}

In \autoref{fig:cluster2} we show the distribution of family members in the three projections of the proper elements space, with the size of the points being proportional to the diameter of the corresponding asteroid. The patterned points are the three asteroids we excluded from the membership as interlopers and are shown only for reference. 

The first thing we notice is the asymmetrical distribution of the small members with respect to the largest one, which also approximates the location of the barycenter of the family.  Indeed as seen clearly in the bottom two panels, all the smaller asteroids have higher inclinations and than (633)~Zelima, and almost all of them have lower eccentricities than it as well. The ellipses in these two panels are derived from the Gaussian equations and represent the outcome of an isotropic ejection resulting in a $\Delta v=12ms^{-1}$ velocity difference from the barycenter.  This is not an unexpected outcome, since cratering collisions can in many cases produce a jet of fragments from the parent body (see e.g. \citet{2011Icar..211..856H}). 

We also notice that four asteroids are separated from the rest of the cluster in semi-major axes, toward slightly lower values. It turns out that these asteroids have chaotic obits as a result of their interaction with the 8J-3S-3 three-body mean motion resonance, a fact that is also reflected by their higher Lyapunov exponents as shown in \autoref{tab:example_table}. 

\begin{figure}
  \includegraphics[width=0.85\columnwidth]{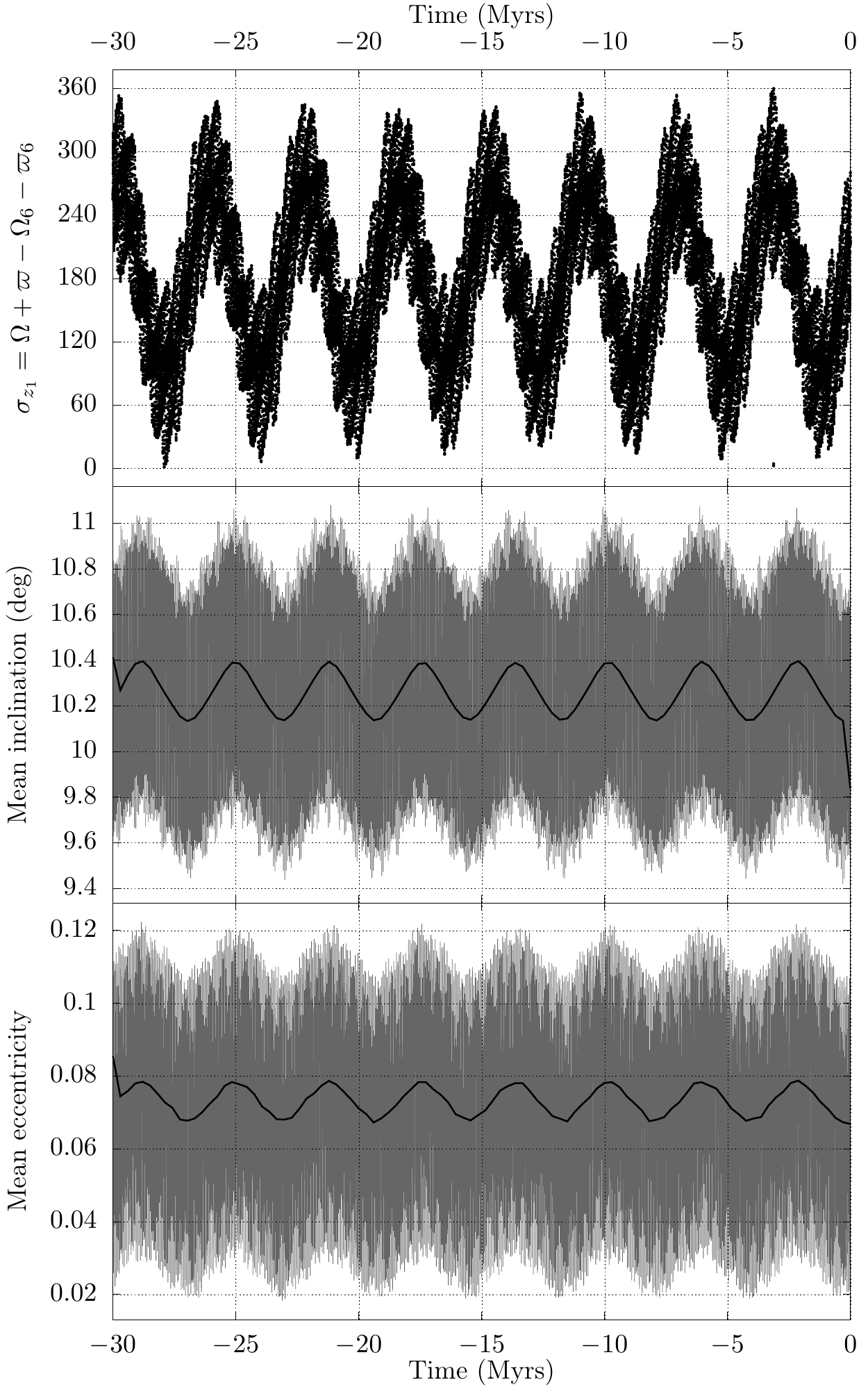}
    \caption{Evolution of the critical argument of $z_1$ ($\sigma_{z_1}=\Omega+\varpi-\Omega_6-\varpi_6$) (top), mean inclination (middle) and mean eccentricity (bottom), of the asteroid (633). The solid black lines in the middle and bottom panels are the respective averaged mean elements, and are shown to better appreciate the effect of the resonance , i.e. the induced oscillations.}
    \label{fig:633ex}
\end{figure}

The region of the (221)~Eos family, is also crossed by the $z_1$ secular resonance, which has been shown \citep{2006Icar..182...92V} to have a significant impact on the orbits of small asteroids, especially in conjunction with the Yarkovsky effect. Having propagated the orbits of the members of the Zelima cluster for the needs of the BIM as mentioned above, we were able to verify that the whole Zelima cluster was born inside this resonance. Constructing the critical argument of the resonance for each asteroid ($\sigma_{z_1}=\Omega+\varpi-\Omega_6-\varpi_6$) as a function of time, we observed large amplitude librations with a period of about 4 Myr. The effect of the resonance on the orbits of family members is the induction of oscillations in their eccentricities and inclinations with amplitudes of the order of 0.005 and 0.12 degrees respectively. An illustrative example can be seen in \autoref{fig:633ex}, where we show the evolution of the critical argument, the mean inclination and mean eccentricity of asteroid (633).

\begin{table}
  \centering
  \caption{Members of the (633)~Zelima cluster,  and their physical properties\protect\footnotemark }
  \label{tab:example_table}
  \begin{tabular}{lccr} 
    \hline
    $Asteroid^a$ & $H^b$  & $D^c$ & $LCE^d$\\
    \hline
    633 & 9.79 &    40.15(w)& 1.54\\
    6733 & 11.81 &  13.05(w) & 1.45\\
    89714 & 14.22 & 4.22(w) &1.16 \\
    119786 & 14.76 &3.76(w) &1.55 \\
    149046 & 14.82 &3.22 &0.40 \\
    230472 & 15.56 &2.30 &2.32 \\
    250011 & 15.34 &2.54 &0.87 \\
    292616 & 15.24 &2.66 &1.24 \\
    298735 & 15.82 &2.04 &1.46 \\
    299846 & 16.12 &1.77 &34.47 \\
    324175 & 15.92 &1.95 &1.27 \\
    364004 & 15.83 &2.03 &1.50 \\
    368472 & 15.79 &2.07 &1.98 \\
    372029 & 15.94 &1.93 &1.69 \\
    402372 & 15.96 &1.91 &1.49 \\
    411116 & 16.70 &1.36 &16.92 \\
    412371 & 16.12 &1.77 &1.76 \\
    421572 & 16.07 &1.82 &41.79 \\
    443574 & 16.65 &1.39 &1.03 \\
    449004 & 16.76 &1.32 &1.40 \\
    456020 & 16.48 &1.50 &1.49 \\
    475448 & 16.76 &1.32 &44.40 \\
    487314 & 16.41 &1.55 &1.12 \\
    \hline
  \end{tabular}
  \begin{tablenotes}
  \item[1]$^a$ Asteroid number
  \item[2]$^b$ Absolute magnitude
  \item[3]$^c$ Diameter in km (available WISE diameters \citep{2011ApJ...741...68M} are marked with (w), otherwise they are calculated from the absolute magnitude assuming a geometric albedo of $p_v=0.2$)
  \item[4]$^d$ Lyapunov characteristic exponent in $Myrs^{-1}$
  \end{tablenotes}
\end{table}

\section{Summary and Discussion}
We have discovered a new sub-family of the very old and large asteroid family of (221)~Eos. By applying the hierarchical clustering method to the catalog of proper elements and using a distance cut-off of $20\,m\,s^{-1}$ we have identified a group of 26 asteroids around the asteroid (633)~Zelima, giving its name to the new cluster. 

We then created 1000 clones of the Entire Eos family and searched within those for any clusterings of 26 or more asteroids  within the same distance cut-off, with no positive results, establishing that the cluster corresponds to the outcome of a real collisional disruption with $>99\%$ statistical certainty. 

Using the backward integration method we were able to identify and remove three interlopers from the family list, based on the fact that their orbits did not converge suitably well with those of the rest of the members. 

In order to get a  precise age estimation, and to eliminate possible errors in the initial conditions and the dynamical model, we run a statistical approach of the backward integration method. For each of the 23 member asteroids we created 100 clones, accounting for the variations in the initial conditions and the Yarkovsky effect parameters. After integrating their orbits we selected $3\cdot10^6$ random combinations of clones of each asteroid and evaluated a function depending on the mean differences in both secular angles, noting the time at which it features a global minimum. Thus we established a statistically accurate age of $2.9\pm0.2$~Myrs for the (633)~Zelima cluster.

A short analysis of the physical and orbital characteristics of the family members led us to the conclusion that the cluster was formed in a cratering event, leaving the parent body with about 98\% of the initial mass and producing a directional jet of fragments leading to the anisotropic distribution of orbital elements we observe today. The dynamical environment of the cluster is in principle stable with only the 8J-3S-3 three-body mean-motion resonance leading four small members to chaotic orbits.

As we mentioned in the Introduction, the Zelima cluster could be of great importance in the study of the Eos family, both dynamically and compositionally. However, the currently available physical data are limited to the diameters and albedos of the few largest members of the cluster (\autoref{tab:example_table}). New data concerning the spectral properties of the cluster members, either from targeted observations or from the next generation of surveys, will enable a deeper study of their physical properties, and in conjunction with the rest of the Eos family members, provide useful insights about the partially differentiated Eos parent body.

Another interesting aspect is the association of recent asteroid disruptions to observed dust bands in the Main Belt \citep{2003ApJ...591..486N,2008ApJ...679L.143N}, and we believe it would be worth trying to find such a link for the Zelima cluster as well.

\section*{Acknowledgements}

I would like to express my gratitude to Bojan Novakovi\'c for his helpful comments concerning the preparation of this manuscript. I would also like to thank the reviewer, David Vokrouhlick{\'y}, for his constructive comments that helped improve the quality of the manuscript.
\bibliographystyle{mnras}
\bibliography{bib}
\end{document}